\documentclass[onecolumn]{revtex4}
\usepackage{amsmath,amssymb}
\usepackage{graphicx}
\usepackage{dcolumn}
\usepackage{epsfig,color}
\usepackage{bm}
\begin{document}

\title{An interacting quark-diquark model. Strange and nonstrange baryon spectroscopy and other observables}

\author{M. De Sanctis}
\affiliation{Universidad Nacional de Colombia, Bogot\'a, Colombia}
\author{J. Ferretti}
\affiliation{Dipartimento di Fisica and INFN, `Sapienza' Universit\`a di Roma, P.le Aldo Moro 5, I-00185 Roma, Italy}
\author{R. Maga\~na Vsevolodovna}
\author{P. Saracco}
\author{E. Santopinto}
\affiliation{INFN, Sezione di Genova, via Dodecaneso 33, I-16146 Genova, Italy}

\begin{abstract}
We describe the relativistic interacting quark-diquark model formalism and its application to the calculation of strange and nonstrange baryon spectra. The results are compared to the existing experimental data. 
We also discuss the application of the model to the calculation of other baryon observables, like baryon magnetic moments, open-flavor strong decays and baryon masses with self-energy corrections.
\end{abstract}

\maketitle

\section{Introduction}
\label{intro}
The diquark concept has been used in a large number of studies, ranging from one-gluon exchange to lattice QCD calculations. For example, see Refs. \cite{Jakob:1997,Jaffe:2003,Wilczek:2004im,Jaffe:2004ph,Santopinto:2004hw,Selem:2006nd}. 

Recently, an interacting quark-diquark model has been introduced by Santopinto in Ref. \cite{Santopinto:2004hw}. This is a nonrelativistic potential model, where baryons are described as two-body quark-diquark bound states and the relative motion between the quark and diquark constituents in terms of a relative coordinate $\vec r$. 
The Hamiltonian contains a Coulomb-like plus linear confining interaction and an exchange one, depending on the spins and isospins of the quark and the diquark.
In this article, we will discuss the main features of the interacting quark-diquark model, as relativistically reformulated in Refs. \cite{Ferretti:2011zz,qD2014b,qD2014a,charmed-baryons} within the point form formalism \cite{Klink:1998zz}. 
 
Finally, we will summarize our results for the strange and nonstrange baryon spectra \cite{Santopinto:2004hw,Ferretti:2011zz,qD2014b,qD2014a} and baryon magnetic moments in the quark-diquark model \cite{qD2014a}. 
We will also briefly discuss the formalism to calculate other baryon observables in the quark-diquark model, including open-flavor strong decays and baryon masses with self-energy corrections.

\section{An interacting quark-diquark model}
\label{Int qD}
In quark-diquark models, baryons are assumed to be composed of a constituent quark and a constituent diquark \cite{GellMann:1964nj,Ida:1966ev,lich}.  
Up to an energy of 2 GeV, the diquark can be described as two correlated quarks with no internal spatial excitations \cite{Santopinto:2004hw,Ferretti:2011zz}. Then, its color-spin-flavor wave function must be antisymmetric. 
Moreover, as we consider only light baryons, made up of $u$, $d$, $s$ quarks, the internal group is restricted to SU$_{\rm sf}$(6). 
If we denote spin by its value, flavor and color by the dimension of the representation, the quark has spin $s_2 = \frac{1}{2}$, flavor $F_2={\bf {3}}$, and color $C_2 = {\bf {3}}$. 
The diquark must transform as ${\bf {\overline{3}}}$ under SU$_{\rm c}$(3), hadrons being color singlets. Then, one only has the symmetric SU$_{\rm sf}$(6) representation $\mbox{{\boldmath{$21$}}}_{\rm sf}$(S), containing $s_1=0$, $F_1={\bf {\overline{3}}}$ (the scalar diquark) and $s_1=1$, $F_1={\bf {6}}$ (the axial-vector diquark) \cite{Wilczek:2004im,Jaffe:2004ph}. 

\subsection{Model Hamiltonian and calculation of the strange and nonstrange baryon spectra}
\label{The Model} 
\begin{figure}[h]
\begin{minipage}{16pc}
\includegraphics[width=18pc]{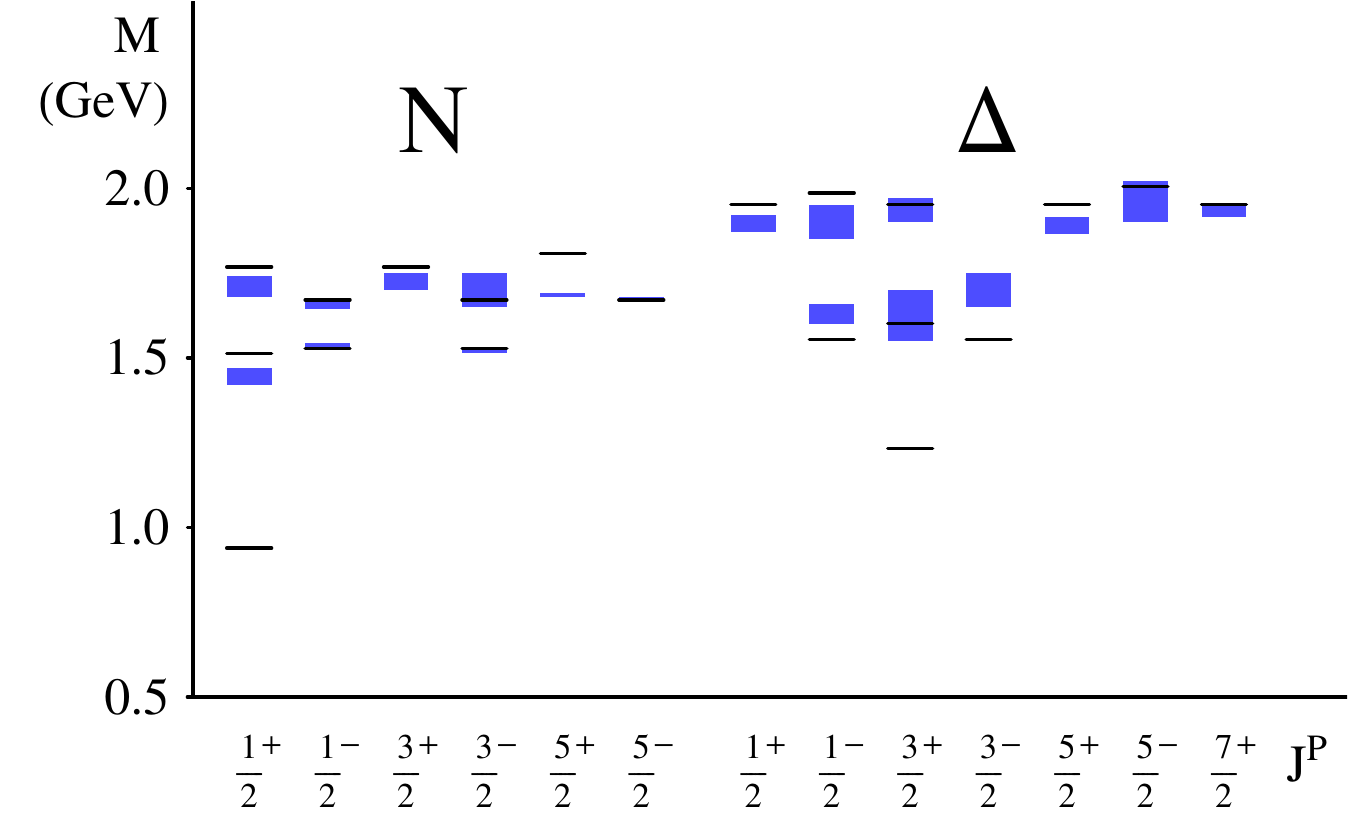}
\caption{\label{fig:Spectrum-ND}Comparison between the calculated masses \cite{Ferretti:2011zz} (black lines) of the $3^*$ and $4^*$ $N$ and $\Delta$ resonances (up to 2 GeV) and the experimental masses from PDG \cite{Nakamura:2010zzi} (blue boxes). APS copyright.}
\end{minipage}
\hspace{2pc}
\begin{minipage}{18pc}
\includegraphics[width=18pc]{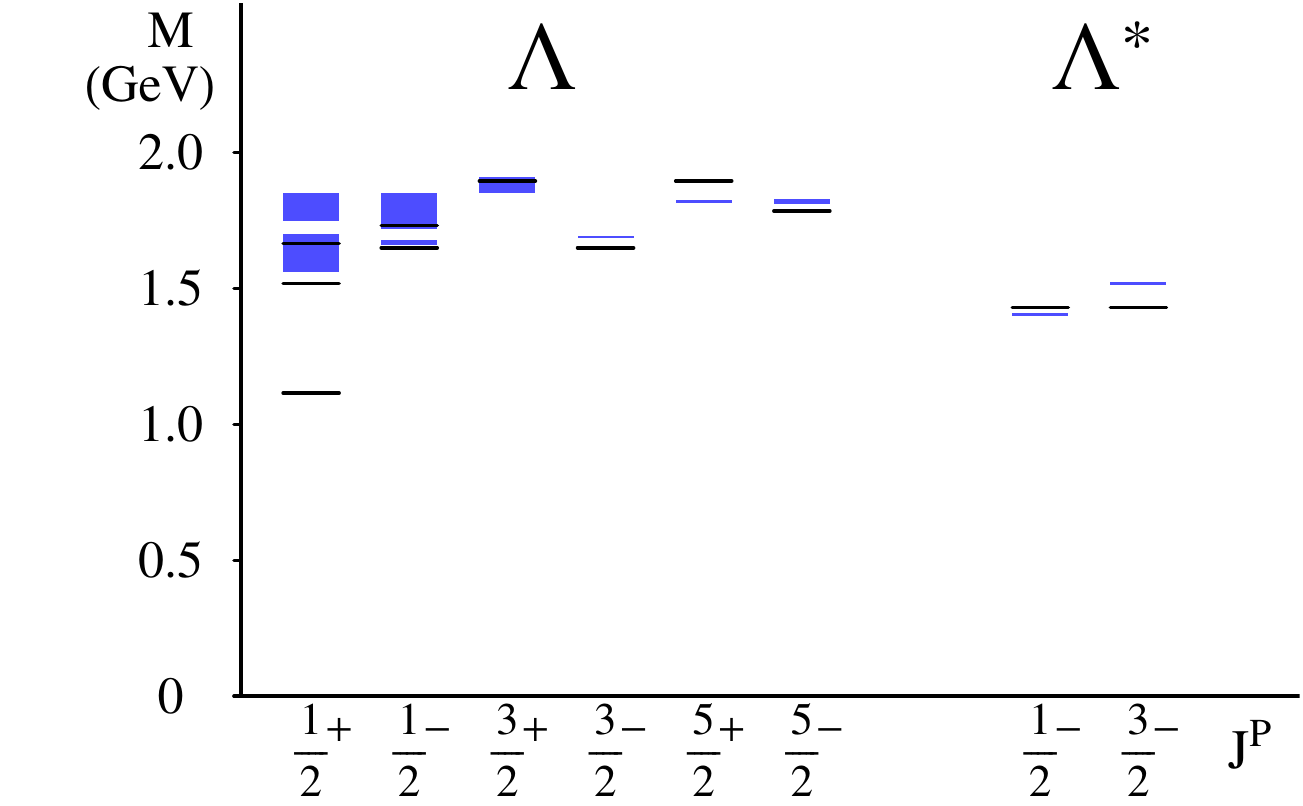}
 \caption{\label{fig:Spectrum-lambda}As Fig. \ref{fig:Spectrum-ND}, but for $\Lambda$ and $\Lambda^*$ resonances. Figure taken from Ref. \cite{qD2014b}; APS copyright.}
\end{minipage}
\end{figure}

\begin{figure}[h]  
\begin{minipage}{16pc}
\end{minipage} \\
\begin{minipage}{16pc}
\includegraphics[width=18pc]{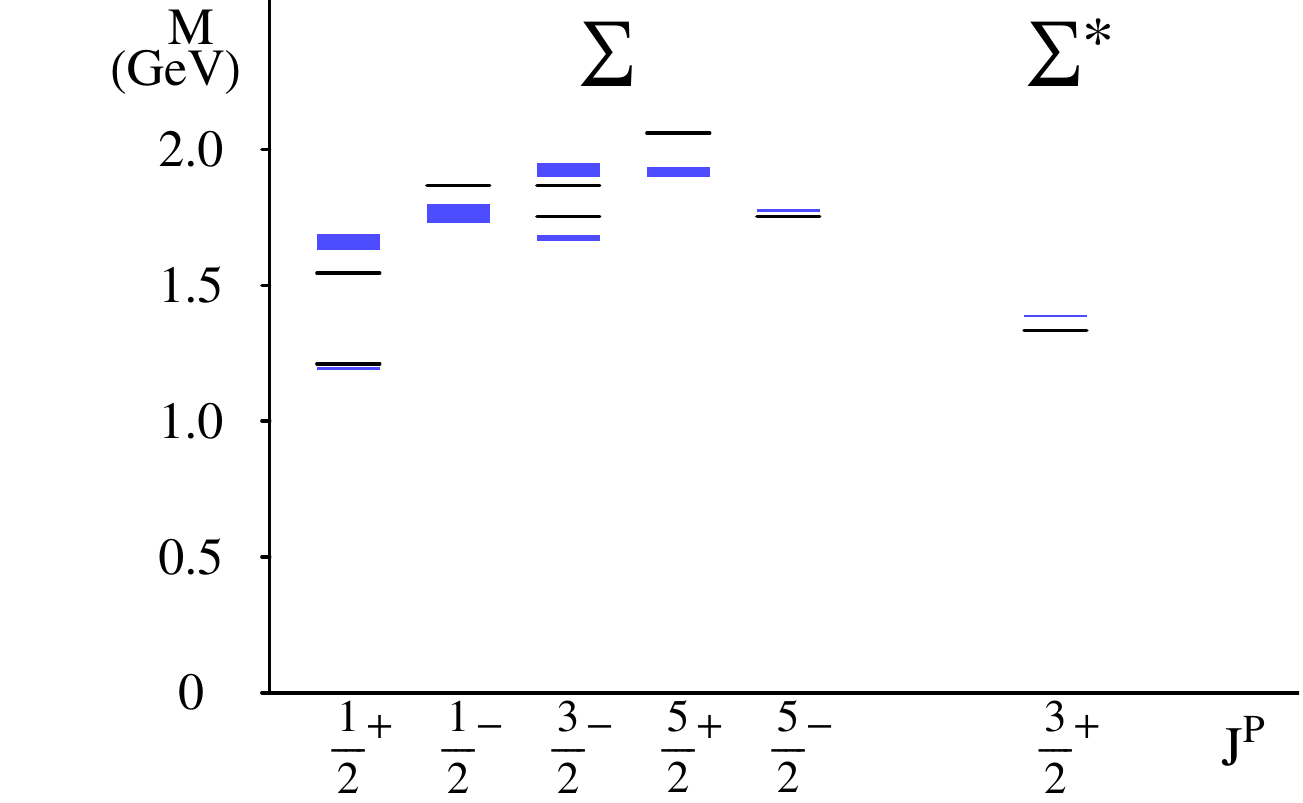}
\caption{\label{fig:Spectrum-sigma}As Fig. \ref{fig:Spectrum-ND}, but for $\Sigma$ and $\Sigma^*$ resonances. Figure taken from Ref. \cite{qD2014b}; APS copyright.}
\end{minipage}
\hspace{2pc}
\begin{minipage}{18pc}
\includegraphics[width=18pc]{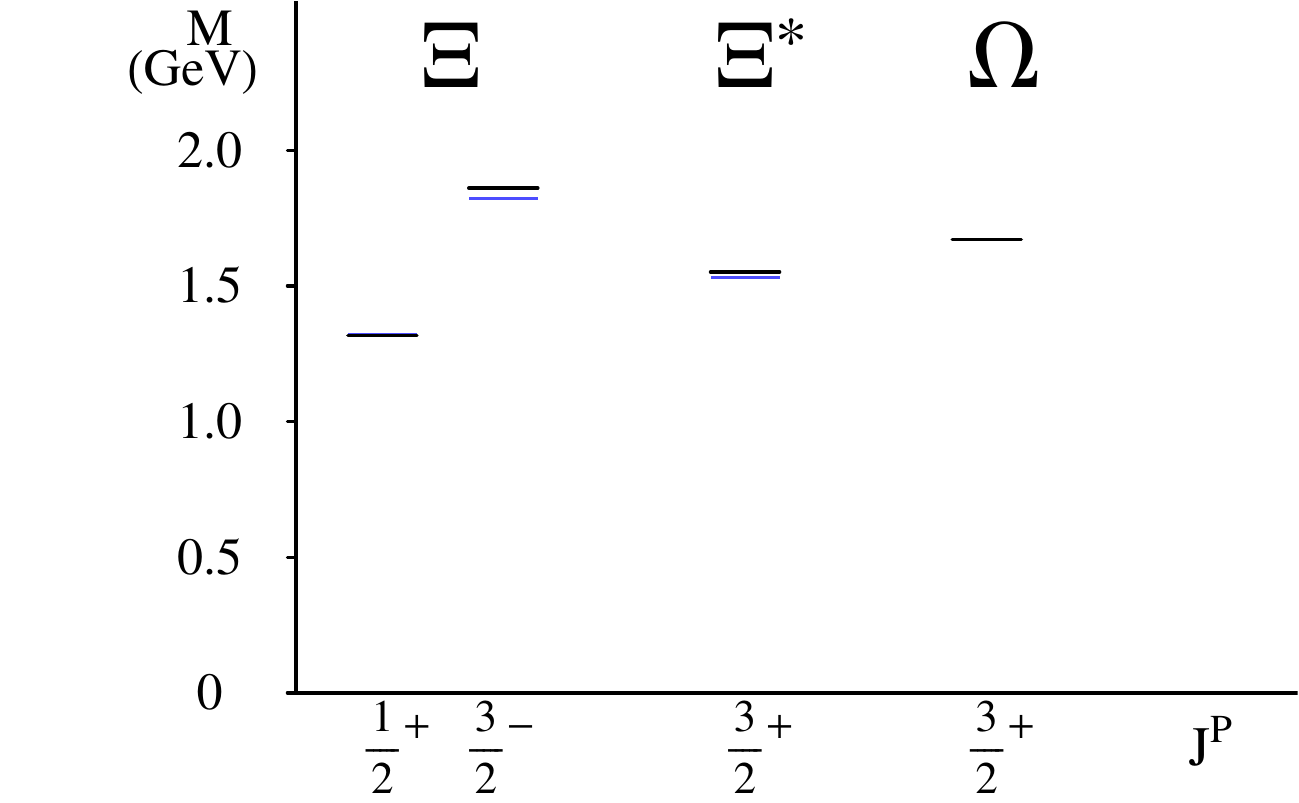}
 \caption{\label{fig:Spectrum-xi-omega}As Fig. \ref{fig:Spectrum-ND}, but for $\Xi$, $\Xi^*$ and $\Omega$ resonances. Figure taken from Ref. \cite{qD2014b}; APS copyright.}
\end{minipage}
\end{figure}

The Hamiltonian of the model is \cite{Ferretti:2011zz}
\begin{equation}
	\begin{array}{rcl}
	M & = & E_0 + \sqrt{\vec q\hspace{0.08cm}^2 + m_1^2} + \sqrt{\vec q\hspace{0.08cm}^2 + m_2^2} 
	+ M_{\rm dir}(r)  + M_{\rm ex}(r) 
	\end{array}  \mbox{ }.
	\label{eqn:H0}
\end{equation}
Here, $E_0$ is a constant, $M_{\rm dir}(r)$ and $M_{\rm ex}(r)$ are the direct and exchange quark-diquark interactions, respectively, $m_1$ and $m_2$ the diquark and quark masses.
The direct term, 
\begin{equation}
  \label{eq:Vdir}
  M_{\rm dir}(r)=-\frac{\tau}{r} \left(1 - e^{-\mu r}\right)+ \beta r ~~,
\end{equation}
is the sum of a Coulomb-like interaction with a cut off plus a linear confinement term. 
The exchange interaction is given by \cite{Santopinto:2004hw,Ferretti:2011zz},
\begin{equation}
	\begin{array}{rcl}
	M_{\rm ex}(r) & = & \left(-1 \right)^{L + 1} \mbox{ } e^{-\sigma r} \left[ A_{\rm S} \mbox{ } \vec{s}_1 
	\cdot \vec{s}_2  + A_{\rm I} \mbox{ } \vec{t}_1 \cdot \vec{t}_2  
	+  A_{\rm SI} \mbox{ } \vec{s}_1 \cdot \vec{s}_2 \mbox{ } \vec{t}_1 \cdot \vec{t}_2  \right]  
	\end{array}  \mbox{ },
	\label{eqn:Vexch-nonstrange}
\end{equation}
where $\vec{s}$ and $\vec{t}$ are the spin and isospin operators. To calculate the strange baryon spectrum, Eq. (\ref{eqn:Vexch-nonstrange}) has to be generalized in the form of a G\"ursey-Radicati interaction \cite{qD2014b,Gursey:1992dc},
\begin{equation}
	\begin{array}{rcl}
	M_{\rm ex}(r) & = & \left(-1 \right)^{L + 1} \mbox{ } e^{-\sigma r} \left[ A_{\rm S} \mbox{ } \vec{s}_1 
	\cdot \vec{s}_2  + A_{\rm F} \mbox{ } \vec{\lambda}_1^{\rm f} \cdot \vec{\lambda}_2^{\rm f} \mbox{ } 
	+ A_{\rm I} \mbox{ } \vec{t}_1 \cdot \vec{t}_2  \right]  
	\end{array}  \mbox{ },
	\label{eqn:Vexch-strange}
\end{equation}
where $\vec{\lambda}^{\rm f}$ are SU$_{\rm {f}}$(3) Gell-Mann matrices. 
In the nonstrange sector, we also have a contact interaction \cite{Santopinto:2004hw,Ferretti:2011zz}, 
\begin{equation}
	\begin{array}{rcl}
	\label{eqn:Vcont}	
	M_{\rm cont}(r) & = & \left(\frac{m_1 m_2}{E_1 E_2}\right)^{1/2+\epsilon} \frac{\eta^3 D}{\pi^{3/2}} 
	e^{-\eta^2 r^2} \mbox{ } \delta_{L,0} \delta_{s_1,1}  \left(\frac{m_1 m_2}{E_1 E_2}\right)^{1/2+\epsilon}
	\end{array}  \mbox{ },
\end{equation}
introduced to reproduce the $\Delta-N$ mass splitting. In Eqs. (\ref{eq:Vdir}-\ref{eqn:Vcont}), $\tau$, $\mu$, $\beta$, $A_{\rm S}$, $A_{\rm I}$, $A_{\rm SI}$, $A_{\rm F}$, $\eta$ and $D$ are free parameters, fitted to the reproduction of the experimental masses \cite{Nakamura:2010zzi}. Our results, extracted from Refs. \cite{Ferretti:2011zz,qD2014b}, are shown in Figs. \ref{fig:Spectrum-ND}-\ref{fig:Spectrum-xi-omega}. These results can be compared to those of three quark models \cite{IK,CI,HC,GR,LMP} and other quark-diquark model calculations \cite{Santopinto:2004hw,Galata:2012xt,Gutierrez:2014qpa,Faustov:2015eba}.

\subsection{Nonstrange baryon spectrum with a spin-isospin transition interaction}
\label{Nonstrange baryon spectrum with a spin-isospin transition interaction}
In Ref. \cite{qD2014a}, we improved the "relativized" model of Ref. \cite{Ferretti:2011zz} by introducing a spin-isospin transition interaction, which induces a mixing between scalar and axial-vector diquark states.
We computed the nonstrange baryon spectrum within point form dynamics and used the resulting wave functions to calculate the magnetic moments of the proton and neutron (see Sec. \ref{Baryon magnetic moments}). 

The spin-isospin transition interaction, $M_{\rm tr}(r)$, was chosen as
\begin{equation}    
	\label{eqn:Vtr(r)}
	M_{\rm tr}(r) = V_0 \mbox{ } e^{-\frac{1}{2} \nu^2 r^2} (\vec s_2 \cdot \vec S) 
	(\vec t_2 \cdot \vec T) \mbox{ },
\end{equation}
where $V_0$ and $\nu$ are free parameters and the matrix elements of the spin transition operator, $\vec S$, are defined as
\begin{equation}
	\left\langle \right. s_1', m_{s_1}' \left. \right| S_\mu^{[1]} \left| s_1, m_{s_1} \right\rangle \neq 0 
	\mbox{ for } s_1' \neq s_1  \mbox{ },
\end{equation}
with $\left\langle 1 \right\| S_1 \left\| 0 \right\rangle = 1$ and $\left\langle 0 \right\| S_1 \left\| 1 \right\rangle = -1$. The matrix elements of the isospin transition operator, $\vec T$, are defined analogously. 
\begin{figure}[htbp] 
\centering 
\includegraphics[width=8cm]{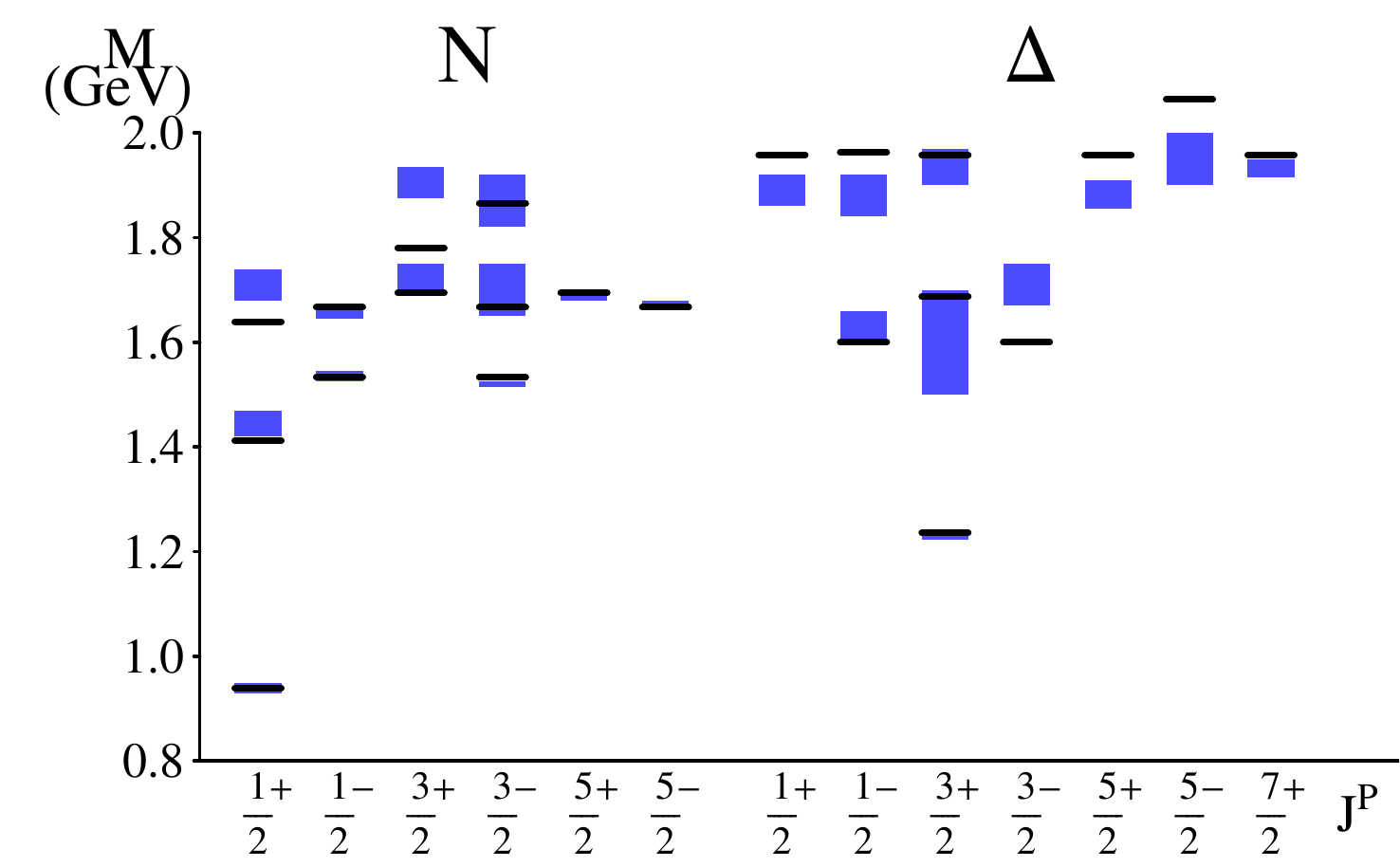}
\caption{Comparison between the calculated masses (black lines) of the $3^*$ and $4^*$ nonstrange baryon resonances (up to 2 GeV) and the experimental masses from PDG \cite{Nakamura:2010zzi} (boxes). Figure taken from Ref. \cite{qD2014a}; Societ\`a Italiana di Fisica/Springer-Verlag copyright.} 
\label{fig:Spectrum3e4}
\end{figure}

The spin-isospin transition interaction of Eq. (\ref{eqn:Vtr(r)}) mixes quark-scalar diquark and quark-axial-vector diquark states, i.e. states with $s_1 = 0$ ($t_1 = 0$) and $s_1 = 1$ ($t_1 = 1$), whose total spin (isospin) is $S = \frac{1}{2}$ ($T = \frac{1}{2}$). 
Thus, in this version of the model the nucleon state, as well as states such as the $D_{13}(1520)$, the $S_{11}(1535)$ and the $P_{11}(1440)$, contain both a $s_1 = 0$ and a $s_1 = 1$ component. 

The introduction of the interaction of Eq. (\ref{eqn:Vtr(r)}) determines an improvement in the overall quality of the reproduction of the experimental data (considering only $3^*$ and $4^*$ resonances) \cite{qD2014a}, with respect to that obtained with the previous version of this model \cite{Ferretti:2011zz}. 
In particular, the Roper resonance, $N(1440)$ $P_{11}$, is far better reproduced than before and the same holds for $N(1680)$ $F_{15}$. See Fig. \ref{fig:Spectrum3e4}.

\section{Baryon observables in the interacting quark-diquark model}
Below, we discuss the formalism to calculate other baryon observables in the interacting quark-diquark model of Ref. \cite{qD2014a}. We stress the importance of having a model in which $N^*$'s wave functions contain both a $q$--scalar diquark and a $q$--axial-vector diquark component, which play a role analogous to that of $\rho$- and $\lambda$-type components in the nucleon wave function. 
We think that the presence of both components is necessary to obtain a good reproduction of baryon observables in a quark-diquark model, including e.g. baryon magnetic moments and open-flavor strong decays.

\subsection{Baryon magnetic moments}
\label{Baryon magnetic moments}
The electromagnetic response of the two-state diquark is in principle unknown.

As a starting point, for the magnetic dipole operator $\vec \mu$ we consider a non-relativistic expression \cite{qD2014a}
\begin{equation}
	\label{eqn:mu-NR}
	\vec \mu = \frac{f}{m} \left(e_1 \vec s_1 + e_2 \vec s_2\right) 
	+ \frac{f_{\rm q}}{m_{\rm q}} \mbox{ } e_{\rm q} \vec s_{\rm q}  \mbox{ }.
\end{equation}
In order to use the permutational symmetry of the quarks 1 and 2 of the diquark, we re-write Eq. (\ref{eqn:mu-NR}) as \cite{qD2014a}:
\begin{equation}
	\label{eqn:mu-eff}
	\vec \mu = \frac{f}{2m} \mbox{ } e_{\rm D} \vec s_{\rm D} + \frac{f}{2m} \mbox{ } \Delta e_{\rm D} \Delta \vec s_{\rm D}  
	+ \frac{f_{\rm q}}{m_{\rm q}} \mbox{ } e_{\rm q} \vec s_{\rm q}  \mbox{ },
\end{equation}
where $e_{\rm D} = e_1 + e_2$ is the charge operator of the diquark in the isospin space, $\vec s_{\rm D} = \vec s_1 + \vec s_2$ represents its spin, $\Delta e_{\rm D} = e_1 - e_2$ is the difference between the charges of quarks 1 and 2 and $\Delta \vec s_{\rm D} = \vec s_1 - \vec s_2$ is the difference between the spin operators of the two quarks. 
Eq. (\ref{eqn:mu-eff}) can be re-written as \cite{qD2014a}
\begin{equation}
	\label{eqn:mu-eff-final}
	\vec \mu = \frac{f_{\rm AV}}{m_{\rm AV}} \mbox{ } e_{\rm D} \vec s_{\rm D}	+ \frac{f_{\Delta}}{\left\langle m 
	\right\rangle} \mbox{ } \Delta e_{\rm D} \Delta \vec s_{\rm D} + \frac{f_{\rm q}}{m_{\rm q}} \mbox{ } e_{\rm q} \vec s_{\rm q}  \mbox{ },
\end{equation}
representing the effective magnetic dipole operator of our model.
Note that the second term gives the $S - AV$ transitions.
The mean values of the effective dipole operator can be easily calculated, so that the numerical values of free parameters ($f_{\rm AV}$ and $f_{\Delta}$) can be fitted to reproduce the proton and neutron magnetic moments, that are $\mu_{\rm p} = 2.793$ n.m.u. and $\mu_{\rm n} = -1.913$ n.m.u. \cite{Nakamura:2010zzi}. 
Because the number of free parameters is larger than that of the experimental informations, we take $f_{\rm q} = 1$; the resulting values for the other parameters are $f_{\rm AV} = 0.0462$ and $f_{\Delta} = 0.146$ \cite{qD2014a}.

A more detailed study of the electromagnetic current will be carried out in a subsequent paper, with the aim of calculating the nucleon electromagnetic form factors and the helicity amplitudes of baryon resonances \cite{DeSanctis:TBP}.

\subsection{Open-flavor strong decays and missing resonances}
As discussed in Refs. \cite{Santopinto:2004hw,Ferretti:2011zz,Galata:2012xt}, the baryon spectra of quark-diquark and three quark models are quite distinct, the main differences being in the amount and energy of theoretically predicted states. When theoretical results and experimental observations are compared, it is clear that three quark models predict an eccessive number of states. This is the problem of missing resonances which can be, at least partially, solved in quark-diquark models in terms of a smaller number of effective degrees of freedom \cite{Santopinto:2004hw,Ferretti:2011zz,Galata:2012xt}. 
Nevertheless, experimental informations on a hadron are not only limited to its mass, but also include (open- and hidden-flavor) strong decay widths, weak and e.m. radiative transitions, and so on. Thus, we think it is also worthwhile to compare three quark and quark-diquark model predictions for some of these observables.

\begin{figure}[htbp] 
\centering 
\includegraphics[width=6cm]{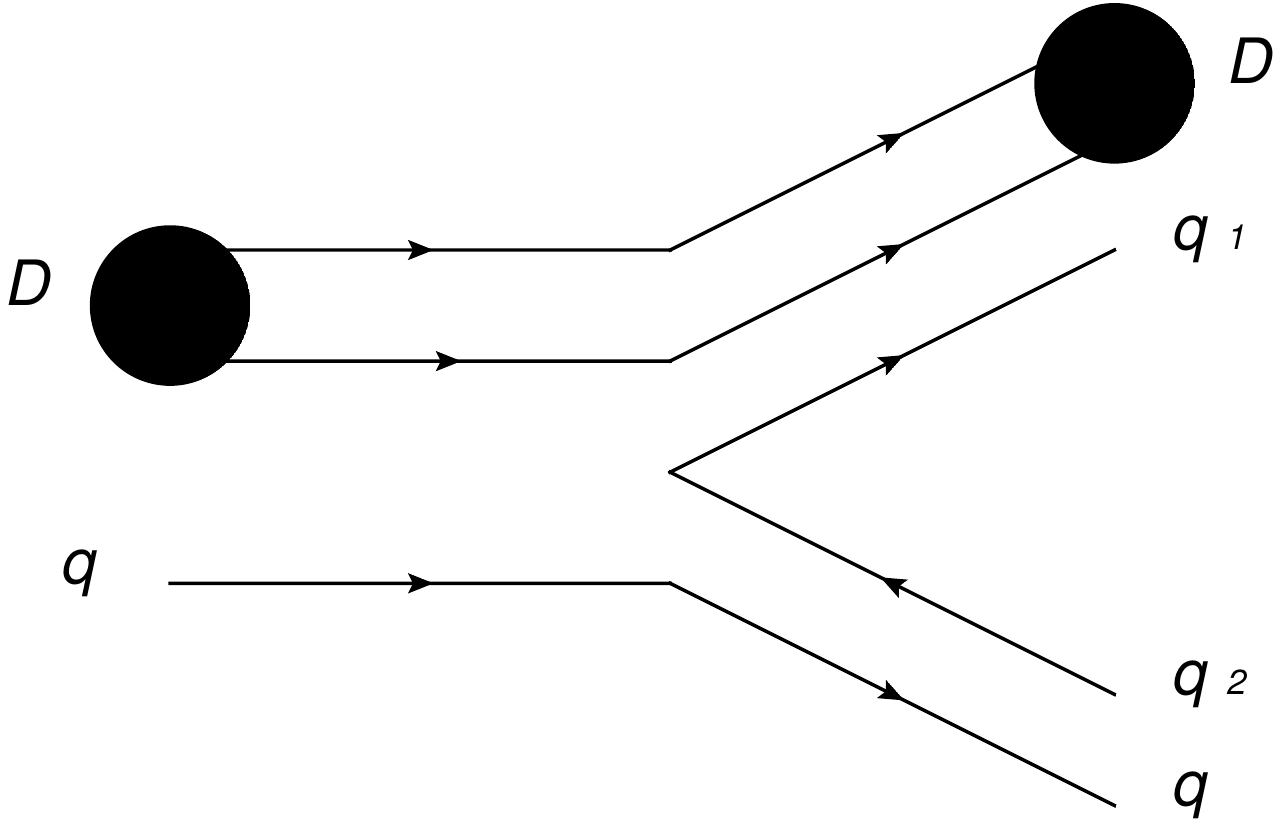}
\caption{The $^3P_0$ pair-creation model of hadron vertices is here used within a quark-diquark picture for baryons. $D$ and $q$ are, respectively, the spectator diquark and quark; $q_1 q_2$ is the created $q \bar q$ pair, with $^3P_0$ quantum numbers. Springer-Verlag Wien copyright.} 
\label{fig:qD-3P0}
\end{figure}

Here, we discuss the formalism to compute the open-flavor strong decays of nonstrange baryons in the well-known $^3P_0$ pair-creation model \cite{Micu,LeYaouanc,Roberts:1992}. 
In the $^3P_0$ model, a hadron decay takes place in its rest frame and proceeds via the creation of an additional $q \bar q$ pair ($q_1$ and $q_2$) with vacuum quantum numbers, i.e. $J^{\rm PC} = 0^{++}$. The decay amplitude $A(qD) \rightarrow B(Dq_1) C(q q_2)$ can be expressed as \cite{Micu,LeYaouanc,Roberts:1992}
\begin{equation}
	\label{eqn:3P0-decays-ABC}
	\Gamma_{A \rightarrow BC} = \Phi_{A \rightarrow BC}(q_0) \sum_{\ell} 
	\left| \left\langle BC q_0  \, \ell J \right| T^\dag \left| A \right\rangle \right|^2 \mbox{ },
\end{equation}
where $\mathcal M_{A \rightarrow BC}(q_0) = \left\langle BC q_0  \, \ell J \right| T^\dag \left| A \right\rangle$ is the so-called $^3P_0$ amplitude, $\Phi_{A \rightarrow BC}(q_0)$ the phase space factor for the decay, $q_0$ and $\ell$ the relative momentum and orbital angular momentum of the $BC$ pair. See Fig. \ref{fig:qD-3P0}. 
We write the phase space factor in the usual relativistic form, $\Phi_{A \rightarrow BC}(q_0) = 2 \pi q_0 \frac{E_{\rm b}(q_0) E_{\rm c}(q_0)}{M_{\rm a}}$, where $E_{\rm i}(q_0) = \sqrt{q_0^2 + M_{\rm i}^2}$ (i = $B, C$).

The transition operator of the $^{3}P_0$ model is given by \cite{Micu,LeYaouanc,Roberts:1992}:
\begin{equation}
	\begin{array}{rcl}
	T^{\dagger} & = & -3 \gamma_0 \, \int d \vec{p}_{\rm q_1} \, d \vec{p}_{\rm q_2} \, \delta(\vec{p}_{\rm q_1} + \vec{p}_{\rm q_2}) \, C_{12} \, F_{12} \, 
	\Psi_{\rm d}(\vec p_{\rm q_1}, \vec p_{\rm q_2}) \left[ \chi_{12} \, \times \, {\cal Y}_{1}(\vec{p}_{\rm q_1} - \vec{p}_{\rm q_2}) \right]^{(0)}_0 \\ 
	& \times & b_1^{\dagger}(\vec{p}_{\rm q_1}) \, d_2^{\dagger}(\vec{p}_{\rm q_2})    
	\mbox{ }.
	\end{array}
\label{3p0}
\end{equation}
Here, $\gamma_0$ is the pair-creation strength, $b_1^{\dagger}(\vec{p}_{\rm q_1})$ and $d_2^{\dagger}(\vec{p}_{\rm q_2})$ the creation operators for a quark and an antiquark with momenta $\vec{p}_{\rm q_1}$ and $\vec{p}_{\rm q_2}$, respectively, $\Psi_{\rm d}(\vec p_{\rm q_1}, \vec p_{\rm q_2})$ the pair-creation vertex or quark form factor. The $q \bar q$ pair is characterized by a color singlet wave function $C_{12}$, a flavor singlet wave function $F_{12}$, a spin triplet wave function $\chi_{12}$ with spin $S=1$ and a solid spherical harmonic ${\cal Y}_{1}(\vec{p}_{\rm q_1} - \vec{p}_{\rm q_2})$, since the quark and antiquark are in a relative $P$ wave. 

\subsection{Baryon spectrum with self-energy corrections in the quark-diquark model}
The formalism described in the previous section can also be used to compute the baryon $\rightarrow$ baryon + meson vertices relevant to the calculation of baryon self-energies. For a calculation of octet and decuplet baryon self-energies within a three quark model, see Ref. \cite{baryon-SE}.

We consider  the Hamiltonian
\begin{eqnarray}
	H = H_0 + V \mbox{ },
\end{eqnarray}
where $H_0$ is the "unperturbed" part, describing the interaction between constituent (valence) quarks [see Eq. (\ref{eqn:H0})], and $V$ the interaction which can couple a baryon state to the baryon-meson continuum.  
The physical mass of a baryon, $M_{\rm a}$, can be written as
\begin{equation}
	\label{eqn:phys-Ma}
	M_{\rm a} = E_{\rm a} + \Sigma(E_{\rm a})  \mbox{ },
\end{equation}
where $E_{\rm a}$ is the bare mass, obtained by solving the eigenvalue problem of $H_0$, and
\begin{equation}
	\label{eqn:self-a}
	\Sigma(E_{\rm a}) = \sum_{BC} \int_0^{\infty} q^2 dq \mbox{ } \frac{\left| V_{\rm a,bc}(q) \right|^2}{E_{\rm a} - E_{\rm bc}}  \mbox{ }
\end{equation}
the self-energy correction.
In Eq. (\ref{eqn:self-a}), one has to sum over a complete set of   baryon-meson  intermediate states, $\left| BC \right\rangle$. These channels, with relative momentum $q$ between $B$ and $C$, have quantum numbers $J_{\rm bc}$ and $\ell$ coupled to the total angular momentum of the initial state $\left| A \right\rangle$. $V_{\rm a,bc}$ stands for the coupling between the intermediate state $\left| BC \right\rangle$ and the unperturbed wave function of the baryon $A$. 
Several choices for $V$ are possible, including the $^3P_0$ pair-creation model operator $T^\dag$ of Eq. (\ref{3p0}), which is ours.

In the following section, we provide an example of a $^3P_0$ model vertex calculation ($\Delta \rightarrow \Delta \pi$) in the quark-diquark model. This particular vertex provides the largest contribution to the $\Delta$ self-energy \cite{baryon-SE}.

\subsection{$\Delta \rightarrow \Delta \pi$ vertex}
As an example, we briefly discuss the calculation of the $\Delta \rightarrow \Delta \pi$ vertex. 
The matrix elements can be calculated according to Refs. \cite{qD2014a,Roberts:1992,Ferretti:2013faa,Strong2015}, with a few differences.
In the special case of $L_{\rm b} = L_{\rm c} = 0$, the transition amplitude can be written as \cite[Sec. 6]{Roberts:1992}
\begin{equation}
	\label{eqn:3P0-general}
	\begin{array}{rcl}
	\mathcal M_{A \rightarrow BC} & = & \frac{1}{3} (6 \gamma_0) \mbox{ } (-1)^{L_{\rm a} + S_{\rm a} + J_{\rm a}} \hat S_{\rm a} \hat S_{\rm b} \hat S_{\rm c} \hat S_{\rm bc} \hat \ell
	\left\{ \begin{array}{ccc} S_{\rm D} & S_{\rm q_1} & S_{\rm b} \\ S_{\rm q} & S_{\rm q_2} & S_{\rm c} \\ S_{\rm a} & 1 & S_{\rm bc}\end{array} \right\}
	\left\{ \begin{array}{ccc} S_{\rm a} & L_{\rm a} & J_{\rm a} \\ \ell & S_{\rm bc} & 1 \end{array} \right\} \\
	& \times & \mathcal F_{A \rightarrow BC}  \mbox{ } \epsilon_{A \rightarrow BC}(q_0)  \mbox{ },
	\end{array}
\end{equation}
where $\mathcal F_{A \rightarrow BC}$ and $\epsilon_{A \rightarrow BC}(q_0)$ are the flavor and spatial matrix elements, respectively, $\vec S_{\rm bc} = \vec S_{\rm b} + \vec S_{\rm c}$ the spin of the $BC$ pair, $S_{\rm D}$ and $S_{\rm q}$ the spin of the spectator diquark and quark, $S_{\rm q_1}$ and $S_{\rm q_2}$ the spins of the $^3P_0$ quark and antiquark. We also use the notation $\hat \ell = \sqrt{2 \ell + 1}$. 
In the $\Delta^{++} \rightarrow \Delta^{++} \pi^0$ case, Eq. (\ref{eqn:3P0-general}) reduces to
\begin{equation}
	\begin{array}{rcl}
	\mathcal M_{\Delta^{++} \rightarrow \Delta^{++} \pi^0} & = & -\frac{\gamma_0}{3 \sqrt 5}  
	\mbox{ } \mathcal F_{\Delta^{++} \rightarrow \Delta^{++} \pi^0}  \mbox{ } \epsilon_{\Delta^{++} \rightarrow \Delta^{++} \pi^0}(q_0)  \mbox{ },
	\end{array}
\end{equation}
The flavor matrix element can be written as \cite{Ferretti:2013faa}
\begin{equation}	
	\label{eqn:flavor-ME}
	\mathcal F_{A \rightarrow BC} = \left\langle BC | A \Phi_0 \right\rangle \mbox{ },
\end{equation}
where $\left| \Phi_0 \right\rangle = \frac{1}{\sqrt 3} \left( \left| u \bar u\right\rangle + \left| s \bar s\right\rangle + \left| s \bar s\right\rangle\right)$ is the SU(3) flavor singlet, $\left| A \right\rangle $, $\left| B \right\rangle $ and $\left| C \right\rangle$ the flavor wave functions of the corresponding hadrons \cite{Wilczek:2004im,Jaffe:2004ph}.
In the $\Delta^{++} \rightarrow \Delta^{++} \pi^0$ case, Eq. (\ref{eqn:flavor-ME}) can be written as
\begin{equation}
	\begin{array}{rcl}
	\mathcal F_{\Delta^{++} \rightarrow \Delta^{++} \pi^0} & = & 
	\left\langle \Delta^{++} \pi^0 | \Delta^{++} \Phi_0 \right\rangle = \frac{1}{\sqrt 6} \left[ \left\langle \{u,u\}u \right| \otimes \left( \left\langle u \bar u \right| 
	- \left\langle d \bar d \right| \right) \right] \\ & & \left[ \left| \{u,u\}u \right\rangle \otimes \left( \left| u \bar u \right\rangle + \left| d \bar d \right\rangle 
	+ \left| s \bar s \right\rangle \right) \right] = \frac{1}{\sqrt 6} \mbox{ },
	\end{array}
\end{equation}
where $\left| \pi^0 \right\rangle = \frac{1}{\sqrt 2}$ $\left( \left| u \bar u \right\rangle - \left| d \bar d \right\rangle \right)$, $\left| \Delta^{++} \right\rangle = \left| \{u,u\}u \right\rangle$, and the notation $\{q,q\}$ stands for an axial-vector diquark \cite{Wilczek:2004im,Jaffe:2004ph}. 

The spatial matrix element are the same as those for meson $\rightarrow$ meson + meson transitions \cite{Roberts:1992,Ferretti:2013faa}. We use the coordinate system $\vec q_{\rm b} = \frac{1}{M_{\rm b}} \left( m_{\rm D} \vec p_{\rm q_1} - m_{\rm q_1} \vec p_{\rm D} \right)$, $\vec q_{\rm c} = \frac{1}{M_{\rm c}} \left( m_{\rm q_2} \vec p_{\rm q} - m_{\rm q} \vec p_{\rm q_2} \right)$, $\vec P_{\rm cm} = \vec P_{\rm b} + \vec P_{\rm c}$, $\vec q = \frac{M_{\rm c}}{M_{\rm bc}} \vec P_{\rm b} - \frac{M_{\rm b}}{M_{\rm bc}} \vec P_{\rm c}$, where $\vec P_{\rm b} = \vec p_{\rm D} + \vec p_{\rm q_1}$, $\vec P_{\rm c} = \vec p_{\rm q} + \vec p_{\rm q_2}$, $M_{\rm b} = m_{\rm D} + m_{\rm q_1}$, and $M_{\rm c} = m_{\rm q} + m_{\rm q_2}$. 
The spatial matrix elements can be written explicitly as
\begin{equation}
	\label{eqn:spatial-general}
	\begin{array}{rcl}
	\epsilon_{A \rightarrow BC}(\vec p_{\rm q}, \vec p_{\rm D}, \vec p_{\rm q_1}, \vec p_{\rm q_2}) & = & \int {\rm d}\vec p_{\rm q} {\rm d}\vec p_{\rm D} {\rm d}\vec p_{\rm q_1} 
	{\rm d}\vec p_{\rm q_2} \mbox{ } \mathcal Y_1(\vec p_{\rm q_1} - \vec p_{\rm q_2}) \delta(\vec p_{\rm q_1} + \vec p_{\rm q_2}) \Psi_{\rm b}^*(\vec p_{\rm D}, \vec p_{\rm q_1})  \\
	& \times & \Psi_{\rm c}^*(\vec p_{\rm q}, \vec p_{\rm q_2}) \Psi_{\rm d}(\vec p_{\rm q_1}, \vec p_{\rm q_2}) \Psi_{\rm a}(\vec p_{\rm q}, \vec p_{\rm D})  \mbox{ },
	\end{array}
\end{equation}
where $\Psi_{\rm a}(\vec p_{\rm q}, \vec p_{\rm D}) = \Psi_{\rm a}^{\rm ho}\left(\frac{1}{M_{\rm a}} (m_{\rm D} \vec p_{\rm q} - m_{\rm q} \vec p_{\rm D})\right) \delta(\vec P_{\rm a} - \vec p_{\rm q} - \vec p_{\rm D})$, $\Psi_{\rm b}(\vec p_{\rm D}, \vec p_{\rm q_1}) = \Psi_{\rm b}^{\rm ho}\left(\vec q_{\rm b}\right) \delta(\vec P_{\rm b} - \vec p_{\rm D} - \vec p_{\rm q_1})$ and $\Psi_{\rm c}(\vec p_{\rm q}, \vec p_{\rm q_2}) = \Psi_{\rm c}^{\rm ho}\left(\vec q_{\rm c}\right) \delta(\vec P_{\rm c} - \vec p_{\rm q} - \vec p_{\rm q_2})$ are single harmonic oscillator (ho) wave functions with ho parameters $\alpha_{\rm a} = \alpha_{\rm b} = 3.14$ GeV$^{-1}$ \cite{qD2014a} and $\alpha_{\rm c} = 2.00$ GeV$^{-1}$ \cite{Ackleh:1996yt}; $\Psi_{\rm d}(\vec p_{\rm q_1}, \vec p_{\rm q_2}) = \mbox{exp}\left[-\frac{\alpha_{\rm d}^2}{6} (\vec p_{\rm q_1} - \vec p_{\rm q_2})^2 \right]$ is a Gaussian quark form factor, with $\alpha_{\rm d} = 1.70$ GeV$^{-1}$ \cite{Ferretti:2013faa}. 
In the case of ground-state hadrons, namely with no radial and orbital excitations, Eq. (\ref{eqn:spatial-general}) reduces to
\begin{equation}
	\label{eqn:spatial-delta++}
	\begin{array}{rcl}
	\epsilon_{\Delta^{++} \rightarrow \Delta^{++} \pi^0}(q_0) & = & - \mathcal N_{\rm a} \mathcal N_{\rm b}^* \mathcal N_{\rm c}^* \frac{x+1}{8 G^3} q_0 e^{-F^2 q_0^2}  \mbox{ },
	\end{array}
\end{equation}
where 
\begin{equation}
	x = - \frac{\frac{m_{\rm D}}{M_{\rm b}} \alpha_{\rm b}^2 + \frac{m_{\rm q}}{M_{\rm c}} \alpha_{\rm c}^2 + \frac{4}{3} \alpha_{\rm d}^2}{\alpha_{\rm a}^2+\alpha_{\rm b}^2+\alpha_{\rm c}^2+\frac{4}{3} \alpha_{\rm d}^2} \mbox{ },
\end{equation}	 
\begin{equation}
	G^2 = \frac{1}{2} \left(\alpha_{\rm a}^2+\alpha_{\rm b}^2+\alpha_{\rm c}^2+\frac{4}{3} \alpha_{\rm d}^2\right) \mbox{ },
\end{equation}	 
\begin{equation}
	F^2 = \frac{1}{2} \left[\alpha_{\rm a}^2 x^2 + \alpha_{\rm b}^2 \left(x + \frac{m_{\rm D}}{M_{\rm b}}\right)^2 + \alpha_{\rm c}^2 \left(x + \frac{m_{\rm q}}{M_{\rm c}}\right)^2 + \frac{4}{3} \alpha_{\rm d}^2 (x+1)^2\right]  \mbox{ },
\end{equation}	
and $\mathcal N_{\rm i} = 2 \pi^{-1/4} \alpha_{\rm i}^{3/2}$ (i = a, b, c).

\end{document}